\documentclass[preprint,showpacs,pra,nofootinbib]{revtex4}
\usepackage{amsmath,amssymb,amsfonts,color,graphicx}

\begin{document} 
\title{Optical activity in the Drude helix model}
\author{Florian Dufey}
\email{dufey@ph.tum.de}
\affiliation{Institut f\"ur Theoretische Physik T38, Technische Universit\"at
  M\"unchen, D-85748
  Garching, Germany}
\date{\today}
\begin{abstract} 
  An old classical one-particle helix model for optical activity, first
  proposed by Drude, is reconsidered here. The quantum Drude model is very
  instructive because the optical activity can be calculated analytically
  without further approximations apart from the Rosenfeld long wavelength
  approximation.  While it was generally believed that this model, when
  treated correctly, is optically inactive, we show that it leads to optical
  activity when the motion of the particle is quantum mechanically treated.
  We also find that optical activity arises even in the classical regime at
  non-zero energy, while for zero energy the model is inactive, in accordance
  with previous results. The model is compared with other one-electron models
  and it is shown that its predicted optical activity is qualitatively
  different from those of other one-electron systems.  The vanishing of
  optical activity in the classical zero-energy limit for the Drude model is
  due to the localization of the particle at the equilibrium position, whereas
  in the analogous model of a particle moving freely on a helix without a
  definite equilibrium position, optical activity does not vanish but the
  spectrum is rescaled. The model under study leads to interesting predictions
  about the optical properties of e.~g.\ helicene derivatives.
\end{abstract} 
\maketitle
\sloppy

\section{Introduction}

Optical activity is of enormous interest in chemistry, e.~g.\ as an analytical
tool to determine the absolute configuration of chiral molecules by comparing
experimentally obtained optical rotatory dispersion spectra with calculated
ones.  Only recently quantum chemical ab initio techniques provide sprectra of
sufficient accuracy for that purpose \cite{Amos82,Pola00,Chee00,Step01}.  On
the other hand, due to their unique optical properties, chiral materials are of
interest on their own: Chiral materials belong to the larger class of
bi-isotropic or, even more generally, bi-anisotropic materials
\cite{Lakh94,Semc96,Serd01}.  Especially interesting in this context
\cite{Rama05} are materials whose electric susceptibility and magnetic
permeability are both negative in a given frequency range.  Meta-materials for
the use in the microwave region consisting of helical structures have been
studied as promising candidates both experimentally
\cite{Lind20,Lind22,Kueh97} and theoretically \cite{Semc89,Semc98,Belo03}. For
applications working in the visible range, one has to pass to a molecular
level. There, quite different classes of molecules with helical structure show
strong optical activity
\cite{Newm56,Sait98,Verb98,Elsh00,Cham04,Jin05,Sanc06}.

For the optimization of these properties, ab initio calculations are only of
limited value as they do not show explicitly the dependence of the properties
on the variables of the system. On the other hand many model systems
\cite{Drud00,Kirk37,Cond37,Tino64,Eyri71,Barr04} have
been devised whose optical activity can be calculated analytically as a
function of their parameters.

In the following, we will be interested in modelling the optical activity of
helical molecules. An especially simple model, which can be solved
analytically, is the one-electron model\footnote{It is important that the
  one-electron models studied in this article can be used to descibe also
  systems of many electrons when these are treated as independent particles.}
for free motion on a helix as proposed by Tinoco and Woody \cite{Tino64}. It
was applied \cite{Leul75} to describe the ORD spectrum of hexahelicene
\cite{Newm56}, a helix shaped molecule built up of six annellated benzene
rings. The model assumed a box shaped potential, which is zero inside the
molecule and infinite outside. However, this implied degeneracy of all
positions inside the molecule is questionable and instable with respect to
perturbations. More recently, helicene derivatives like helicenebisquinones
\cite{Verb98,Elsh00} and tetrathiahelicenes \cite{Cham04} have been shown to
have very large second order non-linear responses making them interesting
materials for non-linear optics. For these helicene derivatives, the
assumption of free motion along the helix cannot be justified any more. Maki
and Persoons \cite{Maki96} instead have assumed that the electronic motion
along the helix is controlled by an effective harmonic potential. This model
probably is historically the first,---it was proposed by Paul Drude
\cite{Drud00} in 1900---, to explain the phenomenon optical activity.
Astonishingly, it has never been treated quantum mechanically, probably
because it was believed that the original classical model is optically
inactive. We will show, however, that this holds true only at zero energy.

\begin{figure}
\begin{center}
\includegraphics{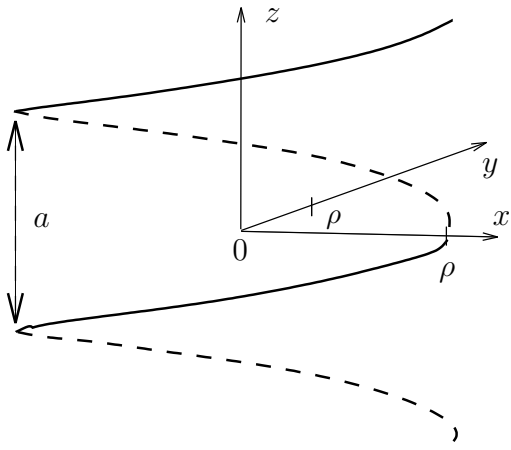}
\end{center}
\caption{\label{fig:Spiral} The spiral considered by Drude with radius $\rho$
  and pitch $a$.}
\end{figure}

Drude considered the motion of a single classical charged particle with charge
$e$ and mass $m_e$ constrained to move on a helical path so that $x(t)=\rho
\cos \phi(t)$, $y(t)=\rho \sin \phi(t)$, and $z(t)=a/(2 \pi)\phi(t)$. Here
$\rho$ is the radius of the helix and $a$ is the pitch (cf.\ fig.\ 
\ref{fig:Spiral}). The helix may be either right or left handed, depending on
the sign of $a$.  The  motion of the particle is governed by the harmonic
Hamiltonian
\begin{equation}
\label{eq:Hamilton}
H(\phi(t),p_\phi(t))=
\frac{1}{2M}p_\phi^2(t)+\frac{M\Omega^2}{2}\phi^2(t),
\end{equation}
 with  $\Omega$ being the frequency. 
The canonical momentum $p_\phi(t)$ is 
\begin{equation}
\label{eq:momentum}
p_{\phi}(t)=\frac{\partial }{\partial \dot
\phi(t)}\left(\frac{1}{2}m_e\mathbf{\dot{x}}^2(t)\right)=M\dot{\phi}(t),
\end{equation}
with the moment of inertia $M=m_e(\rho^2+a^2/(2\pi)^2)$ and the vector
$\mathbf{x}(t)=(x(t),y(t),z(t))^T$.
Drude considered the forced motion of the charged particle under the influence
of the external electric or magnetic fields of a polarized wave and calculated
the resulting electric and magnetic polarization. He found a rotation of the
plane of polarization. In the following it becomes important that Drude
considered forced oscillations, only, and did not take into account any free
oscillations. Hence, the unperturbed system is frozen in at its equilibrium
position and has zero energy.

This model was used extensively in the years following its proposal to
describe optical rotation spectra although soon alternative many particle
models, like the coupled oscillator model, were propagated \cite{Born33}. In
1933, however, Walther Kuhn \cite{Kuhn33} showed that the optical activity of
the helix model was an artifact due to a physically unjustified assumption
made by Drude who took only into account the electric (magnetic) moment
parallel to the axis induced by a magnetic (electric) field component parallel
to the axis, while it neglected the electric (magnetic) moments perpendicular
to the axis induced by a perpendicular magnetic (electric) field.  We will
speak in the following of the parallel and perpendicular component of optical
activity.  So, while Drude only considered the parallel component of optical
activity, Kuhn showed that the parallel and perpendicular components of
optical activities cancel in an isotropic ensemble in the regime of classical
mechanics at zero energy.

Some years afterwards, Condon, Altar and Eyring \cite{Cond37} presented an
alternative quantum mechanical one-particle model which was optically active.
However, they made no attempt to explain why the Drude model did not show
optical activity.  They found that in the classical limit their model also
shows optical activity, but failed to notice that this activity vanishes at
zero energy, too, because it is proportional to the mean quadratic
displacement from the equilibrium position of the particle.

The Drude model was then long forgotten until Desobry and Kabir \cite{Deso73}
showed in the seventies that the Drude model exhibits optical activity in the
nonlinear regime. However, they did not attempt to solve the Drude model
quantum mechanically, neither did they consider the behavior at finite
energies. 

 For the related problem of free motion on a helix, Tinoco and Woody
\cite{Tino64} already established a non-vanishing optical activity in the
quantum regime. While the latter model is formally very similar to the Drude
model, the two models' behavior at low energies is quite distinct as the
particle will become localized at the equilibrium position in the Drude model
while there is no equilibrium position in the case of free motion. 

The outline  of this article is the following:

In sec.\ \ref{sec:QMDrude} we will determine the quantum mechanical expression
for the rotational strength of the Drude model which determines both the
optical rotatory power and the circular dichroism of the model. 

In sec.\ \ref{sec:Limits} we will derive two expansions of the general result:
In the first case, we will expand the optical activity into a series in
$\hbar$ at zero energy. In the second case, we expand into a series in $E$ at
$\hbar=0$, the classical limit. Furthermore, the optical activity of an
oriented sample of helices will be derived. 

In sec.\ \ref{sec:Comparison} we will compare the optical activity of the
Drude model with that of the model of Condon, Altar, and Eyring on the one
hand  and that of the model of free motion on a helix of Tinoco and Woody on the other. 

Finally, a conclusion will be given in sec.\ \ref{sec:Conclusion}.

\section{\label{sec:QMDrude} Rotational strength of the quantum Drude model.}

To determine the optical activity of the Drude helix model
we start from an expression for the rotation $\Phi$ in radians per centimeter
due to Rosenfeld \cite{Rose28}
\begin{equation} \label{eq:Phiquant} \Phi=\frac{8 \pi}{3c \hbar} N_1 \frac{\tilde{n}^2+2}{3} \sum_n \rho_n \sum_m
\frac{\omega^2 R_{mn}}{\omega^2_{mn}-\omega^2}
\end{equation}
where $R_{mn}=\Im\{ \langle n| \mathbf{d}|m\rangle \cdot \langle
m|\mathbf{m}|n\rangle\}$ is the rotational strength and
$\omega_{mn}=(E_m-E_n)/\hbar$ are the eigen-frequencies of the system.  The
particle density is $N_1$, the index of refraction is $\tilde{n}$, while $
\rho_n$ is a statistical weight. The operator of the electric and magnetic
dipole moment, respectively, is $\mathbf{d}=e\,\mathbf{x}$ and
$\mathbf{m}=\frac{e}{2m_ec}\, \mathbf{L}=\frac{e}{2m_ec} \mathbf{x}\times
\mathbf{p}$. The problem to determine of the optical activity thus reduces to
the evaluation of the rotational strength $R_{mn}$.  Addition of an
infinitesimal imaginary part $\omega \to \omega +i \epsilon$ to the frequency
in expression \ref{eq:Phiquant} removes the singularities at
$\omega=\omega_{mn}$ and at the same time, the rotation $\Phi$ acquires an
imaginary part upon this substitution which can be shown to correspond to the
circular dichroism of the transition from state $n$ to state $m$. Using the
well known formula $\lim_{\epsilon \to
  0}(x+i\epsilon)^{-1}=\mathrm{P}(x^{-1})-i\pi \delta (x)$, where
$\mathrm{P}(x^{-1})$ means the principal part, it is easy to show that the
circular dichroism spectrum consists of lines centered at the frequencies
$\omega_{mn}$.

The position of a particle on a helix is described by only one parameter
$\phi$.  While in classical mechanics it is straightforward to find the
corresponding momentum $p_\phi$, in quantum mechanics \cite{Powe69,Bala76}
one has to specify much more carefully how the (approximate) restriction to
the one-dimensional sub-manifold,---in our case a helix---, is achieved before
the corresponding operators $\phi$, $p_\phi$ and their commutation relations
can be found. For the special case of the motion on a helix, there exists a
very simple recipe \cite{Tino64} for quantization: it can be shown that the
operator $p_\phi$ fulfills a canonical commutation relation with $\phi$,
$[p_\phi,\phi]=\frac{\hbar}{i}$, whence $p_\phi=
\frac{\hbar}{i}\frac{\partial}{\partial \phi}$ so that the hamiltonian
\ref{eq:Hamilton} is easily quantized.

 The position operator is formally identical to the classical expression,
\begin{equation} 
\mathbf{x} =\rho \left(\cos\phi, \sin\phi,\frac{a}{2\pi \rho}\phi\right)^T.
\end{equation} 
The cartesian momentum operator $\mathbf{p}$ can be expressed by the time
derivative of $\mathbf{x}$, $\mathbf{p}= m_e \mathbf{\Dot{x}}$. As the
general expression for the time derivative of an operator $A$ depending only
on $\phi$ but not on $p_\phi$ is $\dot{A}=\frac{i}{\hbar}[H,A]=1/2
\{\dot{\phi},\frac{\partial A}{\partial \phi} \}$, we may derive an explicit
expression for $\mathbf{p}$ and finally for the angular momentum operator
$\mathbf{L}=\mathbf{x}\times \mathbf{p}$.  Actually, it can be shown that in
case of the Drude model the operator $\mathbf{L}$ itself may be expressed as
the time derivative of an ``angle'' vector $\mathbf{W}$ which depends on
$\phi$, only, $\mathbf{L}=\frac{i}{\hbar} [H,W]\frac{m_e\rho a}{2\pi}$. The
explicit expression for $\mathbf{W}$ being
\begin{equation}
\label{eq:W}
\mathbf{W}= \left(-2\cos \phi -\phi\sin\phi
    ,-2\sin \phi +\phi \cos\phi,
    \frac{2\pi\rho}{a}\phi \right)^T.
\end{equation}
Hence, the rotational strength becomes
\begin{align}
R_{mn}&=\frac{e^2}{2m_ec}\frac{m_e\rho a}{2\pi}\Im\{\langle
n|\mathbf{x}|m\rangle\cdot \langle
m|\frac{i}{\hbar}[H,\mathbf{W}]|n\rangle\}\notag
\\&=\omega_{mn}\frac{ e^2 \rho a}{8\pi c}
\Re\{\langle n|\mathbf{x}|m\rangle \cdot \langle m|\mathbf{W}|n\rangle\}\notag
\\&=\omega_{mn}\frac{ e^2 \rho a}{16\pi c}\{\langle n|\mathbf{x}|m\rangle \cdot \langle
m|\mathbf{W}|n\rangle+\langle n|\mathbf{W}|m\rangle \cdot \langle
m|\mathbf{x}|n\rangle\},
\end{align}
 where we have assumed the functions   
$\psi_{n,m}(\phi)$ to be real valued in the last line. 

Introducing the vector 
\begin{equation}
\label{eq:Ktilde}
\mathbf{K}=\left(\gamma^{-2}\cos \gamma \phi, \gamma^{-2}\sin \gamma \phi,
    \gamma \phi\right)^T,
\end{equation}
we may write the rotational strength in a more symmetrical form
\begin{equation}R_{mn}=\omega_{mn}\frac{ e^2 \rho^2 a}{8\pi c}\left(\frac{\partial}{\partial \gamma} |\langle
n|K|m\rangle|^2\right)_{\gamma=1}\end{equation}
which is easily shown to hold true. 
This expression can be simplified further to give
\begin{multline}
\label{eq:Rgeneral}
R_{mn}= \omega_{mn}\frac{ e^2\rho^2 a}{8\pi c}\times\\ \times \left(\frac{\partial}{\partial \gamma}
\left(\gamma^{-4} |\langle
n|e^{i\gamma \phi}|m\rangle|^2+|\langle
n|\gamma \phi |m\rangle|^2\right)_{\gamma=1}
\right).
\end{multline}

Up to now, the form of the functions $\psi_{n/m}(\phi)$ or the dependence of
$\omega_{mn}$ on $m$ and $n$ was not of relevance, whence we can use formula
\ref{eq:Rgeneral} to calculate not only the optical activity of the Drude
model but also e.~g.\ that for free motion on a helix with given winding
number which should allow to reproduce the results of Tinoco and Woody
\cite{Tino64}.  In case of the Drude model the states of the system are
characterized by the single quantum number $n$ so that $E_n=\hbar n\Omega$.
The appearing matrix elements are well known \cite{Koid60},
\begin{equation} 
\label{eq:Lagmat}
\langle
n|e^{i\gamma \phi}|m\rangle=e^{\lambda^2/2}\sqrt{\frac{N!}{(N+K)!}}
  \lambda^K L_N^K(-\lambda^2)
\end{equation}
with $\lambda=i\gamma \sqrt{\hbar/(2M\Omega)}$, $N=\mathrm{min}(m,n)$, and
$K=|m-n|$. The $L_N^K$ are Laguerre polynomials,
\begin{equation}
L_N^K(x)=\sum_{j=0}^K  \binom{N+K}{N-j}\frac{(-x)^j}{j!}.
\end{equation} 
Also, $|\langle n|\gamma \phi |m\rangle|^2=\gamma^2\frac{\hbar}{2M\Omega}(N+K)
\delta_{K,1}$ with the Kr\"oneker symbol $\delta_{i,j}=1$ for $i=j$ and
$\delta_{i,j}=0$ for $i\ne j$.

Finally, we find the following expression for the rotational strength of the
Drude model,
\begin{multline}\label{eq:Rquant} R_{mn}=\omega_{mn}\frac{e^2\rho^2 
  a}{8\pi c}\Bigg(\frac{\partial}{\partial \gamma}
\Bigg(\gamma^{-4}\left(\gamma^2\frac{\hbar}{2M\Omega} \right)^K
  e^{-\gamma^2\frac{\hbar}{M\Omega}}\frac{N!}{(N+K)!}\left(L_N^K\left(\gamma^2\frac{\hbar}{2M\Omega}\right)\right)^2+\\ +\gamma^2\frac{\hbar}{2M\Omega} (N+K)\delta_{K,1}\Bigg)_{\gamma=1}\Bigg).
\end{multline}

Exploiting the relation $\frac{\partial}{\partial x}
L_N^K(x)=(-1)L_{N-1}^{K+1}(x)$, the derivation with respect to $\gamma$ may
easily be carried out.

\section{\label{sec:Limits} Classical and low energy limits of the
  optical activity and optical activity of oriented helices.}

We will now consider two limiting cases of that general expression. The first
one is the classical limit. The second limit is the low energy limit in which
only the ground state ($n=0$) is occupied.  The combined classical and $E=0$
limit was considered by Drude and Kuhn and it is interesting to explore why
and how the optical activity tends to zero with $\hbar$ and $E$.

In the classical limit, the quantum number $n$ tends to infinity as $\hbar$
tends to 0 while the energy $E$ is hold fixed.  Radiative transitions will
only be observed between states with quantum numbers $n$ and $m$ such that $K$
is a small number. Hence, $\Delta E$ tends to zero and energy becomes a
continuous function of time $E=E_n \approx E_m$ to order
$\mathrm{O}(\hbar^0)$. The classical motion will be periodic with period
$\Omega$.  Furthermore, the observed transition frequencies will become
multiples of this fundamental frequency, $\omega_{mn}=\pm K\Omega$, as is well
known from the correspondence principle \cite{Pers50}.  We will eliminate
$\hbar$ in favor of $N$ and $E$ from eq.\ \ref{eq:Lagmat}, $\hbar=E/(N\Omega)$.

In this limit (cf.\ ref.\ \cite{Abra72}, eq.\ 22.15.2) the Laguerre
polynomials approach Bessel functions, $\lim_{N \to \infty} N^{-K}
L_N^K\left({x}/{N}\right)=x^{-K/2} J_K(2\sqrt{x})$.  To apply the last formula
to our problem we write $-\lambda^2=\gamma^2\frac{E}{2M\Omega^2}/N$ so that finally
\begin{multline}
\label{eq:Rclass}
R_{mn}=\frac{e^2\rho^2 a K \Omega}{8 \pi c}
\frac{\partial}{\partial \gamma} \Bigg(\gamma^{-4} J^2_K\Bigg(2\gamma
    \sqrt{\frac{E}{2M\Omega^2}}\Bigg)+\\+\gamma^2
  \frac{E}{2M\Omega^2}\delta_{K,1}\Bigg)+\mathrm{O}(\hbar).
\end{multline}

In the following, we will call $R_{mn}$ alternatively $R_K(E,\Omega)$.  We are
still not done in deriving the classical limit of eq.\ \ref{eq:Phiquant} as
this equation contains a pre-factor $\hbar^{-1}$.  In the classical limit,
always a large number of quantum states is occupied so that we have to specify
the function $\rho_n$. E.~g.\ we may assume a micro-canonical distribution in
which only states in a small energy range from $E$ to $E+dE$ are occupied.
For all $n=E/(\hbar \Omega)$ in that range we have $\rho_n=\hbar \Omega/dE$.
If both states $n$ and $m$ are inside the interval $dE$ then, as
$R_{m,n}=-R_{n,m}$, the contribution of these states will cancel. Trivially,
if both $n$ and $m$ are outside the interval, $R_{mn}=0$, too. Only the $K$
states with $n$ inside the interval and $m$ outside (or vice versa) will make
a non-vanishing contribution so that
\begin{multline}
\sum_n \sum_m \frac{\rho_n R_{mn}\omega^2}{\omega_{mn}^2-\omega^2} =\\ \rho_N
\sum_K \frac{K\omega^2}{K^2\Omega^2-\omega^2}
\{R_K(E+dE,\Omega)-R_K(E,\Omega)\}=\\\hbar
\Omega \sum_K \frac{K\omega^2}{K^2\Omega^2-\omega^2} \frac{\partial
  R_K(E,\Omega)}{\partial E},
\end{multline}
whence the classical analog of the Rosenfeld formula becomes 
\begin{equation} 
\label{eq:Rclass2}
 \Phi^\text{class}=\frac{8 \pi }{3c} N_1 \frac{\tilde{n}^2+2}{3} 
\sum_K\frac{\omega^2}{K^2\Omega^2-\omega^2} K \Omega\frac{\partial
  R_K(E,\Omega)}{\partial E}.
\end{equation}
This formula will not only hold true in case of the harmonic oscillator. In
general, $\Omega$ will then be a function of $E$, too.

Expression \ref{eq:Rclass} depends on $E$ only in the combination
$E/(M\Omega^2)=\langle \phi^2 \rangle$, which is the mean quadratic
fluctuation of the angle $\phi$.  For small values of $E$, we may use the
Taylor expansion of the Bessel function
to find the expansion of   $R_K(E)$ into a series in $E$. 
The first non-vanishing term in that series is of order $\mathrm{O}(E^2)$,
\begin{equation}
\label{eq:Phiclamicro}
\Phi^\text{class}=\frac{ e^2 \rho^2 a N_1 }{3Mc^2} \frac{\tilde{n}^2+2}{3} \left(\frac{E}{2M\Omega^2}\right)^2
\sum_K\frac{K^2\omega^2}{K^2\Omega^2-\omega^2} \left(\frac{5}{4}\delta_{K,1}-\frac{1}{2}\delta_{K,2}+\frac{1}{12}\delta_{K,3}\right)+\mathrm{O}(E^3),
\end{equation}
correspondingly, there are three lines in the circular dichroism spectrum at
$\omega=\Omega$, $2\Omega$, and $3\Omega$.

In experiments, probably the only distribution that may be realizable in an
ensemble is the canonical distribution $\rho_n=\exp(-\beta E_n)/\{\sum_m
\exp(-\beta E_m)\}$ with $\beta=1/(kT)$ where $T$ is the temperature.  In the
classical limit, where $E_n$ becomes continuous, the transition from the
micro-canonical optical activity \ref{eq:Phiclamicro} to the canonical
corresponds to a Laplace transform which specializes to the replacement of
$E^2$ in eq.\ \ref{eq:Phiclamicro} by $2(kT)^2$. We see that the optical
activity will thaw up proportional to $T^2$

The second interesting limiting case is that of zero energy when all the
systems are in their ground states. The corresponding expression for the
rotational strength $R_{0m}$ is considerably simpler than the general
expression \ref{eq:Rquant}, as $L_0^K(x)=1$. We may expand this result into a
series in $\hbar$ to find the lowest order correction to the classical optical
activity from which we already know that it vanishes. We find that also the
first order correction in $\hbar$ vanishes, too. The second order result reads
\begin{equation}
\label{eq:REnull}
\frac{R_{m0}}{\hbar}=\frac{e^2\rho^2a}{4\pi c\hbar}\left(\frac{\hbar}{2M \Omega}\right)^3\left(2\delta_{K,1}-\delta_{K,2}+\frac{1}{6}\delta_{K,3}\right)+\mathrm{O}(\hbar^3).
\end{equation}
Again, the circular dichroism spectrum will consist in lowest order of three
lines centered at the fundamental frequency $\Omega$ and its first and second harmonic.

We may compare this result to the prediction that would arise if we would
start like Drude from the physically unjustified approximation of considering
the parallel component of optical activity, $R_{mn}\approx
R_{mn}^\parallel=\Im\{ \langle n| d_z|m\rangle \langle
m|m_z|n\rangle\}=\omega_{mn}\frac{e^2}{4mc}\frac{ m\rho^2
  a}{2\pi}\frac{\partial}{\partial \gamma}(|\langle n|\gamma \phi
|m\rangle|^2|)_{\gamma=1}$, only. The correct result, eq.\ \ref{eq:Rgeneral},
can be seen to be a sum of two terms, $R_{mn}=R_{mn}^\perp +R_{mn}^\parallel$
with $R_{mn}^\perp =\omega_{mn}\frac{e^2}{4mc}\frac{ m\rho^2
  a}{2\pi}\frac{\partial}{\partial \gamma} (\gamma^{-4} |\langle n|e^{i\gamma
  \phi}|m\rangle|^2)|_{\gamma=1}$.  Hence, the optical rotation in the Drude
approximation is
\begin{equation} 
\label{eq:PhiDrude} 
\Phi_\parallel=\frac{e^2\rho^2 a}{3c^2M} N_1 \frac{\tilde{n}^2+2}{3} 
\frac{\omega^2}{\Omega^2-\omega^2}.
\end{equation}
This expression for the optical activity does not depend on $\hbar$ nor on
$E$, furthermore, the response of the system would always be linear for any
strength of the driving field.

We infer that at zero energy and in the linear and classical regime, the
parallel and the perpendicular components of optical activities cancel. When
the extension of the wavefunctions increases, either due to quantum effects,
due to non-zero kinetic energy of the unperturbed system, or when the field
strength is increased, the perpendicular response will become an-harmonic and a
net optical rotation will arise.

Kauzman \cite{Kauz57} noted that a two electron system in which the electrons
are correlated so that their position $\phi$ differs always by $180^\circ$ will
show a specific rotation as predicted by eq.\ \ref{eq:PhiDrude}. It is not
difficult to convince oneself that in this model the perpendicular responses
of the two electrons mutually cancel.

These considerations lead naturally to the consideration of the optical
activity of an oriented sample of helices.  A nice example of such a system
which has been studied experimentally \cite{Verb98,Elsh00} is a
Langmuir-Blodgett film composed of supramolecular arrays of helicene
derivatives.

  As a sample of parallel oriented
helices will be an uniaxial system, optical rotation will only be observed, if
the propagation vector of light is parallel to the axis of the
helices.
In this case both the electric and magnetic field will be perpendicular to the
propagation vector of light, so that one might expect that the optical
activity is determined by the component $R_{mn}^\perp$. In fact, this is not
entirely true, as there are also contributions \cite{Barr04} which are due to
interference of electric dipole transitions with electric quadrupole
transitions so that
\begin{multline}
\label{eq:Rz}
R^\text{axial}_{mn}=\\\frac{\omega_{mn}}{3c}\Re\{\langle m|d_y|m\rangle \langle
  m|\Theta_{xz}|n \rangle -\langle m|d_x|m\rangle \langle
  m|\Theta_{yz}|n \rangle\} + \\ +\Im \{\langle m|d_x|m\rangle \langle
  m|m_x|n \rangle +\langle m|d_y|m\rangle \langle
  m|m_y|n \rangle\}.
\end{multline}
Here, $d_{x,y}$ and $m_{x,y}$ are the $x$- and $y$-components, respectively,
of the electric and magnetic dipole moment vector $\mathbf{d}$ and
$\mathbf{m}$ while $\Theta_{xz}=\frac{3e}{2} xz$ and $\Theta_{yz}=\frac{3e}{2}
yz$ are components of the tensor of the electric quadrupole moment.  In the
second line of eq.\ \ref{eq:Rz} we recognize the contribution $R_{mn}^\perp$,
already defined. The additional contribution due to the quadrupole moments is
easily calculated. As a final result we find 
\begin{equation}
R^\text{axial}_{mn}=\omega_{mn}\frac{
  a}{2\pi c}\{|\langle n|d_x|m\rangle|^2+|\langle n|d_y|m\rangle|^2\}.
\end{equation}
On the other hand, the absorption of the sample due to electric dipole
transitions is proportional to $\omega_{mn}f^\text{axial}_{mn}$ with the
oscillator strength $f^\text{axial}_{mn}=\frac{4\pi^2m_e}{e^2\hbar}\{|\langle
n|d_x|m\rangle|^2+|\langle n|d_y|m\rangle|^2\}$. It is astonishing that e.~g.\
the circular dichroism and the absorption are directly proportional to each
other with the proportionality constant being a measure of the chirality of
the helices. Furthermore, this result is independent of the potential seen by
the charged particle, whence it holds both for the Drude model and for the
model of free motion on a helix. It would be interesting to test this
prediction for the helicene systems mentioned above \cite{Verb98,Elsh00}
for which the classical Drude model has already proven usefull for the
calculation of the second-order optical response \cite{Maki96}.  We also
note that the optical activity will not vanish for an oriented sample, not
even in the classical limit at $E=0$.

\section{\label{sec:Comparison} Comparison with other one-particle models}

We would like to compare the Drude model with two other one-particle models
for optical activity, the model of Condon, Altar, and Eyring on the one hand
side and the model of free motion on a helix with finite length on the other.
As the quantum mechanical expression for arbitrary quantum numbers is complex
on the one hand but lacks specific features on the other we will concentrate
in our analysis on the classical and low energy limit.

The model of Condon, Altar, and Eyring assumes that the particle moves in a
potential of the form $V=\frac{1}{2}(k_1x_1^2+k^2x_2^2+k_3x_3^2)+Ax_1x_2x_3$
with he $k_i$ are force constants and $A$ controls the strength of the
cubic anharmonicity. Condon et al.\ did solve this model quantum mechanically
treating the cubic anharmonicity as a perturbation. They found that the
optical activity of the system is proportional to the quadratic amplitudes of
oscillation $\langle x_i^2 \rangle$ along the main axes of the unperturbed
harmonic oscillator. In the classical limit these amplitudes will tend
linearly to zero when the energy goes to zero.  As any analytic potential may
be expanded into a Taylor series of the assumed form around the equilibrium
position and as in the classical limit at small energies the motion of the
particle will explore only a small region around the equilibrium position, the
response of generic classical systems at low energies will be given by
the result derived by Condon et al.

The vanishing of the optical activity as $E \to 0$ is easily understood as the
amplitudes of oscillation $\langle |x_i| \rangle$ are proportional to $
\sqrt{E}$. At sufficiently small energies the motion of the particle will be
nearly harmonic with the influence of the cubic anharmonicity, which is
responsible for the optical activity, becoming less and less important as
energy is lowered. This conclusion will not only hold true for isotropic
samples but also for oriented systems as for quantum motion within a potential
quadratic in Cartesian co-ordinates, electric and magnetic dipole (or electric
quadrupole) transitions are never allowed simultaneously between any two
states.  This line of argumentation clearly will not hold for the motion being
restricted to the helical path as in the case of the Drude model as the
potential is not analytic.  Although in the Drude model the motion along the
helix is harmonic, electric and magnetic dipole transitions are possible at
the same time between two given states as the co-ordinate $\phi$ does
parameterize a curved path.

To understand the vanishing of the optical activity in the classical limit in
case of the Drude model we expand $\exp(i\gamma \phi)$, which appears in eq.\ 
\ref{eq:Rgeneral}, into a Taylor series.  At zero energy we are left with
matrix elements $\langle M|\phi^k|0 \rangle \sim \hbar^{k/2}$, whence to
lowest order in $\hbar$ we find
\begin{equation}
\gamma^{-4}|\langle M| \exp(i \gamma \phi)|0 \rangle|^2 \sim \gamma^{-2}
|\langle M|  \phi|0 \rangle|^2+\mathrm{O}(\hbar^2)
\end{equation}
so that the first and second term in eq.\ \ref{eq:Rgeneral} cancel. As in case
of the model of Condon et al., the vanishing of optical activity is due to the
strong localization of the position of the particle in the classical low
energy limit. We saw already in the preceeding section when discussing the
original prediction of Drude that an ensemble of oriented helices will show
optical activity even in the classical zero-energy limit.  Furthermore, we
remark that, as in the semiclassical low energy limit, the extension of the
particle's wavefunction will always tend to zero, the use of the Rosenfeld
long wavelength approximation will automatically be justified.The free motion
of a charged particle on a helix of finite extension is quite different in
character from the two models just discussed, as, even in the combined
classical and low energy limit, the particle will not become localized as a
unique equilibrium position is lacking. Hence, we expect that optical activity
will not vanish in the classical limit at $E=0$.  The expression for the
optical activity is more complicated than that of the Drude model, as circular
dichroism is not only to be expected at multiples of one frequency $\Omega$
but at all frequencies $\omega_{mn}=\frac{\hbar
  \pi^2}{2M\phi^2_{\text{max}}}(m^2-n^2)$ with $n,m \in [1,\infty]$ and
$0<\phi<\phi_{\text{max}}$.  The rotatory power of this model was calculated
by Tinoco and Woody \cite{Tino64}.  The energy corresponding to the level $n$
being $E_n=\hbar^2 n^2\pi^2/(2M\phi_\text{max}^2)$, we obtain the classical
limit substituting $m^2+n^2= 16Mt^2E/\hbar^2$ and
$m^2-n^2=\frac{8Mt^2}{\hbar}K\Omega(E)$ (again, $K=m-n$) with
$\Omega(E)=\sqrt{E}/(2t\sqrt{2M})$ and dropping terms of higher order than
$\mathrm{O}(\hbar^0)$. Here, $t=\phi_\text{max}/2\pi$ is the winding number of
the helix.  The resulting expression for the classical rotatory power is of
the form $R_{mn}=MK\Omega f(E/(MK^2\Omega^2))$.  If we insert this into eq.\ 
\ref{eq:Rclass2} we notice that the appearing combination $K \Omega(\partial
R_K(E,\Omega))/(\partial E)$ is actually independent of $E$. The only energy
dependence enters through the frequency factor
$\omega^2/(K^2\Omega^2-\omega^2)$. Hence, a change in energy will only scale
the spectrum or, to put it differently, the optical rotatory dispersion and
circular dichroism spectra will be invariant if $\omega$ is measured in units
of $\Omega$.  The vanishing of the optical activity in the models of Drude and
Condon et al.\ is not paralleled by the model of free motion on a helix.

\section{\label{sec:Conclusion} Conclusion}

In sec.\ \ref{sec:QMDrude} we derived a quantum mechanical expression for the
optical rotatory strength of the Drude helix model. The key result of this
calculation is eq.\ \ref{eq:Rquant}. We found that the Drude model is optical
active in general, with exception of the classical limit at zero energy where
the optical activity vanishes. Incidentally, this was just the limit analyzed
by Drude \cite{Drud00} and Kuhn \cite{Kuhn33}.  To understand better
the reason why optical activity vanishes in that limit, in sec.\
\ref{sec:Limits} we found an expression for the classical optical activity as
a function of $E$ (eq.\ \ref{eq:Rclass2}). We found that optical activity
thaws up like $E^2$ in the classical limit, a behavior that could have been
predicted in principle already by Drude or Kuhn.  Similarly, an expansion of
the general result in powers of $\hbar$ for $E=0$ revealed that there are
non-vanishing quantum corrections of order $\hbar^2$ to the classical
behavior.

While in the general case, the circular dichroism spectrum will consist of
lines at any multiple of the fundamental frequency $\Omega$, in both limits
$E\to 0$ at $\hbar=0$ and $\hbar \to 0$ at $E=0$, respectively, only lines at
frequencies $\omega=n\Omega$ with $n\in [1,2,3]$ will be observed in order
$E^2$ and $\hbar^2$, respectively. We compared this result with the original
(and wrong) prediction due to Drude. He considered only part of the response
of the helix to the external applied field, namely the components parallel to
the axis of the helix, while neglecting the perpendicular component. We
observed that the parallel component of optical activity is independent of
both $\hbar$ and $E$ and gives rise to a line in the circular dichroism
spectrum at the fundamental frequency $\Omega$. On the other hand, the
perpendicular component is strongly dependent on $E$ and $\hbar$ and will be
anharmonic in general. Only in the classical limit and at $E=0$ the parallel
and perpendicular component of optical activity will be of like magnitude but
opposite sign, so that the total optical activity vanishes.

We also consider the possibility to study separately the contribution of these
two components by taking into account an oriented sample of Drude helices as
to be found e.~g.\ in Langmuir Blodgett films of helicene derivatives
\cite{Maki96,Verb98,Elsh00}. These oriented samples form a bi-anisotropic
system. The only case where optical rotation will be observed occurs when the
light propagates parallel to the optical axis, which coincides with the axes
of the helices. We predicted that the optical activity will not vanish in that
case even in the classical limit at $E=0$. Instead, the optical activity will
become proportional to the oscillator strength describing electric dipole
transitions. Thus we conclude that the vanishing of the isotropic optical
activity in the classical limit at $E=0$ is due to a delicate balance between
the individual optical activities for different orientations of the axes.

In sec.\ \ref{sec:Comparison} we compared the Drude model with two other
models describing one electron optical activity, namely the models of Condon,
Altar, and Eyring \cite{Cond37} on the one hand  and the model of Tinoco and
Woody \cite{Tino64} on the other. While all three models are optically active,
they show qualitatively different behavior, especially in the combined
classical and low energy limit. In that limit, the model of Condon et al.\ is
generical for systems of one electron moving in an analytical potential with
unique equilibrium position. We found that the vanishing of optical activity
in that limit is also due to the localization of the particle at the
equilibrium position where it does not perceive the intrinsic chirality of the
potential. However, as this model is truly three dimensional, the optical
activity vanishes proportional to $\hbar$ and $E$ while in the Drude model,
with its one-dimensional potential which is not an analytic function of
$\mathbf{x}$, optical activity vanishes like $\hbar^2$ and $E^2$.  Finally,
the model of Tinoco and Woody, which describes the free motion of a particle
on a helix, is qualitatively different from the other two models considered as
there is no designated equilibrium position. We found that due to this
difference in the classical limit the spectrum does not vanish as a function
of $E$ but remains invariant if the frequency is measured in units of the
natural frequency of the system, $\Omega(E)$, which itself is proportional to
$\sqrt{E}$.


\end{document}